\documentclass{article}
\usepackage{spconf,amsmath,graphicx,pifont,booktabs}
\newcommand{\cmark}{\ding{51}}%
\newcommand{\xmark}{\ding{55}}%

\title{Low-Resource Self-Supervised Learning with SSL-Enhanced TTS}

\name{
  \begin{tabular}{@{}c@{}}
    Po-chun Hsu$^1$, Ali Elkahky$^2$\textsuperscript{*}, Wei-Ning Hsu$^3$, Yossi Adi$^{3,4}$, Tu Anh Nguyen$^3$, Jade Copet$^3$, \\Emmanuel Dupoux$^3$, Hung-yi Lee$^1$, Abdelrahman Mohamed$^5$\textsuperscript{*}\thanks{*Work done while at Meta.}
  \end{tabular}
}
\address{
  $^1$Graduate Institute of Communication Engineering, National Taiwan 
  University\\
  $^2$Apple\\
  $^3$Meta AI\\
  $^4$The Hebrew University of Jerusalem\\  
  $^5$Rembrand Inc.
}

\begin{document}
\maketitle
\vspace{-0pt}
\begin{abstract}
Self-supervised learning (SSL) techniques have achieved remarkable results in various speech processing tasks. Nonetheless, a significant challenge remains in reducing the reliance on vast amounts of speech data for pre-training. This paper proposes to address this challenge by leveraging synthetic speech to augment a low-resource pre-training corpus. We construct a high-quality text-to-speech (TTS) system with limited resources using SSL features and generate a large synthetic corpus for pre-training. Experimental results demonstrate that our proposed approach effectively reduces the demand for speech data by 90\% with only slight performance degradation. To the best of our knowledge, this is the first work aiming to enhance low-resource self-supervised learning in speech processing.
\end{abstract}

\vspace{-0pt}
\section{Introduction}
\label{sec:int}
Using enormous amounts of unlabeled data and tiny fractions of labeled ones helped self-supervised learning (SSL) methods achieve remarkable results in the field of computer vision~\cite{chen2021exploring, grill2020bootstrap}, natural language processing~\cite{brown2020language, lewis2020bart}, and speech processing~\cite{hsu2021hubert, chen2022wavlm}.

Such dependency, however, has raised concerns for certain applications~\cite{huang2022large, pan2020privacy, chen2021evaluating}, may be incorporated at pre-training, leading to potential leakage to downstream applications. 
In the field of speech processing, the leakage can be related to speaker identity or unique timbre in speech utterances~\cite{Tseng2021MembershipIA}. The text content, on the other hand, is simpler to anonymize by using only public domain text data~\cite{panayotov2015librispeech}; 
hence, it doesn't face the same level of scrutiny as speech data that deeply entangles speaker identity and style into the spoken content. 

\begin{figure}[t]
  \vspace{-0pt}
  \centering
  \includegraphics[width=.95\linewidth]{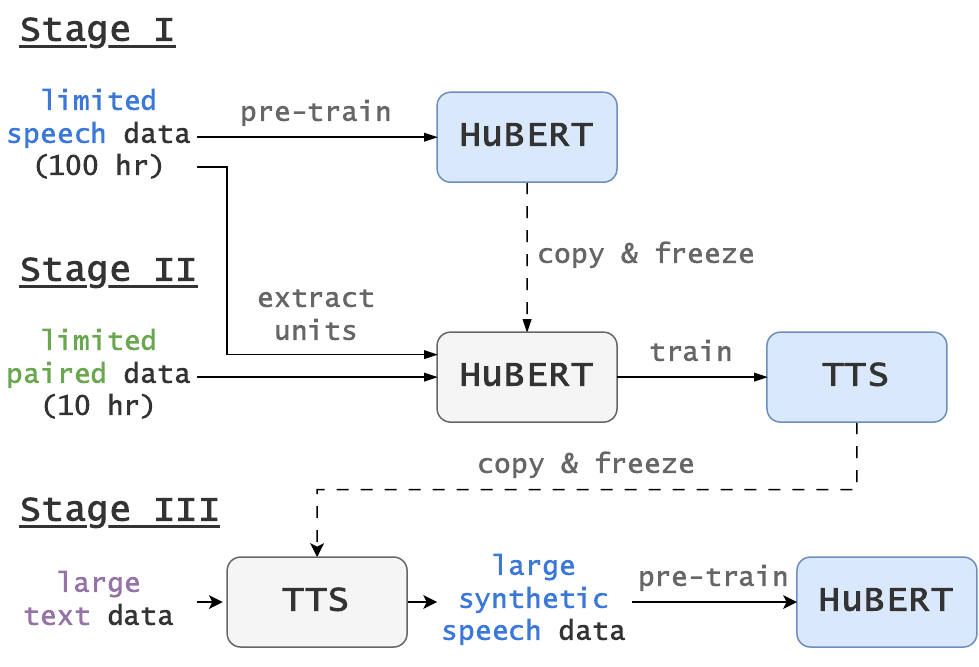}
  \caption{Overview of the proposed system.}
  \label{fig:overview}
  \vspace{-0pt}
\end{figure}

Previous work proposed leveraging additional text data in speech SSL model pre-training~\cite{chen2021injecting, zhang2022speechlm}, and adopted synthetic data to improve ASR performance~\cite{fazel21_interspeech, giollo21_interspeech}. Furthermore, \cite{berrebbi2022more} analyzes the effects of the amount and the number of speakers' data on SSL performance. 
However, none of the TTS systems used in these approaches is build with the presented noisy and ultra-low resources. 
Also, none of these works aim to develop an SSL model with a restricted amount of unpaired speech data. To the best of our knowledge, this is the first work constructing a high-quality SSL model utilizing low-resource speech data.

This work aims to alleviate the need for large amounts of real speech data required for training SSL models by leveraging synthetic data to augment the training corpus. For example, can we train a competitive SSL speech model solely using 100 hours of audio-only data and 10 hours of paired audio and labeled data? Our approach utilizes the high-quality discrete representation units learned by the SSL model to train a unit-to-speech model in an unsupervised fashion. Initially, an SSL model is pre-trained on a small amount of real speech data. Discrete units are then extracted from the speech data using the SSL model for constructing a multi-speaker unsupervised unit-to-speech model. We build a high-quality TTS system by learning text-to-unit mapping using only a limited text-speech paired data. The following phase involves building a large synthetic corpus to pre-train a better SSL model. 
The contributions of this work are summarized as follows:
\begin{itemize}
    \item We investigate the performance of a speech SSL model pre-trained solely on a 100 hours audio-only dataset and demonstrate the impact of data scarcity on SSL models, including the challenge of overfitting and the resulting degradation in performance due to the limited pre-training data.
    \item We propose a TTS system to augment the limited data for SSL pre-training. By leveraging discrete units obtained from the SSL model pre-trained on 100 hours of audio-only data, we build unit-to-speech and text-to-unit models to construct a high-quality multi-speaker TTS system, which is then used to generate a large pre-training speech corpus.
    \item Experimental results show that the proposed method effectively reduces the demand for real-world pre-training data by 90\% with only slight performance degradation compared to SOTA SSL approaches on the standard Librispeech benchmark.
\end{itemize}


\vspace{-0pt}
\section{Proposed System}
\label{sec:pro}

\vspace{-0pt}
\subsection{Overview}
In this work, we choose HuBERT~\cite{hsu2021hubert} as the SSL model to improve in a low-resource setting. The training datasets are in three settings: limited text-speech paired data, limited speech data, and rich-resource text data. More information about the datasets is detailed in Section~\ref{ssec:datasets}.

The high level idea of this work is illustrated in Fig.~\ref{fig:overview}. In \textbf{Stage I}, we first pre-train a HuBERT model on the limited unpaired speech data. This model is used in \textbf{Stage II} to extract discrete speech representation \cite{hsu2021hubert}. We use the SSL representation to help build a multi-speaker TTS system and generate a large synthetic dataset from the rich-resource text dataset. The synthetic dataset is then augmented and used for pre-training a better HuBERT model (\textbf{Stage III}).

\vspace{-0pt}
\subsection{Extracting Discrete Speech Representation}
\label{ssec:s2u}
In \textbf{Stage II}, we first extract speech representations using a HuBERT model. The model is a 12-layered BASE model~\cite{hsu2021hubert} and pre-trained for three iterations on the limited speech dataset in our setting. We then extract the representations from the $9^{th}$ layer and apply k-means clustering to convert continuous features into discrete units. We set $k=500$ in our experiments. Lastly, we adopt the duration-penalized dynamic programming (DPDP) algorithm proposed in \cite{kamper2022word} and remove repeated tokens~\cite{hsu2021hubert}. The duration penalty is set to 1.0.

Table~\ref{tab:s2u-ratio} shows the average length ratios between unit and phoneme sequences with different postprocessing methods. The DPDP algorithm smooths the unit sequence, resulting in a shorter sequence after removing repetitions, which we found helpful for building the TTS system.
\begin{table}[t]
  \vspace{-0pt}
  \caption{Average length ratios (unit length / phoneme length) with different postprocessing methods.}
  \label{tab:s2u-ratio}
  \vspace{-0pt}
  \centering
  \resizebox{0.3\textwidth}{!}{
  \begin{tabular}{ccc}
  \toprule
  \textbf{Remove repetitions} & \textbf{DPDP} & \textbf{Ratio} \\ \hline
  \xmark & \xmark & 4.2 \\ 
  \cmark & \xmark & 2.1 \\ 
  \cmark & \cmark & 1.7 \\ \bottomrule 
  \end{tabular}}
  \vspace{-0pt}
\end{table}

\vspace{-0pt}
\subsection{Text-to-speech Modules}
A common multi-speaker TTS pipeline usually consists of an acoustic feature predictor and a vocoder to restore waveform from acoustic features~\cite{shen2018natural, ren2020fastspeech}. The former is usually trained on a large amount of paired data, and the latter only on speech data. In our setting, the total speech data for training is limited to about 100 hours, and the text-speech paired data is even scarcer (about 10 hours). With only 10 hours of paired data, it is hence challenging to build a multi-speaker TTS from scratch. Some low-resource TTS works leverage additional data from rich resource languages~\cite{xu2020lrspeech} or apply ASR methods to help TTS training~\cite{liu2020towards, ren2019almost}, while in this work, we convert speech into discrete representation to reduce the difficulty of training.

The proposed TTS system comprises four components: (1) A \textit{session encoder} (SE) encodes each speech utterance into a vector, representing information such as speaking style in the utterance. The representation is used during training and generation. (2) A \textit{text-to-unit} model (T2U) predicts the discrete units representing speech information from given text. In preliminary experiments, we found that it was significantly easier to predict discrete units than continuous acoustic features such as Mel-spectrograms. With only a limited amount of paired data, a multi-speaker text-to-Mel model failed to converge and failed to generate intelligible speech. (3) Two variance predictors predict more detailed prosody and pitch information from discrete units. A \textit{duration predictor} (DP) predicts the duration of each discrete unit, and a \textit{pitch predictor} (PP) predicts the $log(F0)$ at each frame. It is worth noting that only speech data is required to train both predictors. Compared with a common TTS model, which takes paired data to train the whole acoustic feature predictor~\cite{shen2018natural, ren2020fastspeech}, the proposed method allows training parts of the acoustic feature predictors without text labels; hence, the amount of training data is less restricted. (4) Finally, a \textit{unit-to-speech} model (U2S) is trained to synthesize the waveform using predicted discrete units and acoustic features.

\vspace{-0pt}
\subsubsection{Session Encoder (SE)}
We follow \cite{snyder2018x} to build SE and extract x-vectors as session embedding. \footnote{Session encoder training is done on NTU infrastructure on publicly available data.}
The model is trained on the limited speech dataset in our setting. To preserve the diversity of speech utterances, we do not average the x-vectors of the same speaker. Instead, we extract an x-vector for each speech utterance and use the utterance-level x-vectors for training and synthesis.

\vspace{-0pt}
\subsubsection{Text-to-unit Model (T2U)}
T2U is implemented based on Tacotron 2~\cite{shen2018natural}. A text sequence is first converted into a phoneme sequence and passed as input to T2U, which then generates a unit sequence. The session embedding from SE is concatenated to the output of the encoder, allowing T2U to generate units in different speaking styles. We append end-of-sentence (EOS) tokens to the end of both input and output sequences, guiding the decoder to predict a stop token when attending to the last input or an EOS token has been predicted at the previous timestep. We remove repeated tokens in target unit sequences during training, and the model follow \cite{wang2017tacotron} to predict two units at each timestep. This process shortens the output length and makes it closer to the input length, leading to faster convergence speed and more stable attention alignments, as mentioned in \cite{wang2017tacotron}.

\vspace{-0pt}
\subsubsection{Duration Predictor (DP) and Pitch Predictor (PP)}
Similarly to~\cite{kreuk2022textless, maimon2022speaking} a duration predictor is trained to restore repeated tokens in unit sequences removed in training T2U. The model takes a deduplicated unit sequence as input and predicts the number of repetitions for each unit.

To generate more acoustic information before restoring the speech waveform, we train a pitch predictor. The model takes a unit sequence (with repetitions) as input and predicts the log-scaled fundamental frequency (F0) at each frame.

Both predictors adopt the same architecture. The input sequence is first encoded by convolution layers with skip connections, followed by an LSTM layer and a combination of several convolution layers and layer normalization. The session embedding is concatenated to each timestep of the input. The predictors are trained to minimize the mean absolute error, and $log(F0)$ is used for calculating the loss.

\vspace{-0pt}
\subsubsection{Unit-to-speech Model (U2S)}
We follow \cite{polyak2021speech,kreuk2022textless} to build a unit-to-speech model with some modifications. The original unit-based HifiGAN from \cite{polyak2021speech} uses the normalized and quantized $log(F0)$ to provide additional pitch information, while in this work, we simply use the unnormalized and continuous $log(F0)$. U2S takes in units, $log(F0)$, and the x-vector to reconstruct the waveform.

To further enhance the quality of the generated speech, we add a variational auto-encoder (VAE) to model the acoustic information that is not explicitly captured in units, $log(F0)$, and x-vectors. During training, a Mel-spectrogram is first extracted from the target waveform, and the VAE encoder encodes the Mel-spectrogram into sequences of mean and variance vectors, which are then used to generate latent representations by the reparameterization trick. The latent representations, the units, the $log(F0)$, and the x-vector are concatenated and passed to the decoder to reconstruct the original Mel-spectrogram. The output of the last hidden layer is used as an additional condition for U2S. At inference time, we discard the encoder and use the normal Gaussian distribution as the prior of the latent representations. The VAE is built using convolution layers with skip connections, and we train the VAE and U2S jointly.

\vspace{-0pt}
\subsubsection{Data Augmentaion and Oversampling}
To enrich the diversity of the synthetic dataset, we augment the generated utterances with two methods. First, we observed that the generated utterances exhibit a length 80\% shorter than natural speech. Consequently, for each utterance, we multiply the durations predicted by DP with a scalar uniformly sampled between 1.0 to 1.5. This process stretches the speech, making the speaking rate various and closer to natural speech. Secondly, we add background noise from our noise dataset to the generated speech with a random SNR between 0 to 15. 

Finally, we combine the augmented synthetic data and the limited real-world speech data as the training set to build the HuBERT model. We found that oversampling the natural speech utterances improves the performance. Specifically, with an oversampling rate of $r$, the natural utterances are sampled for pre-training $r$ times more frequently than synthetic ones.

\vspace{-0pt}
\section{Experimental Setup}
\label{sec:exp}

\vspace{-0pt}
\subsection{Datasets}
\label{ssec:datasets}
We use LibriSpeech~\cite{panayotov2015librispeech} for pre-training, fine-tuning, and evaluation. The dataset is split into several splits for different purposes. All the speech data is downsampled to 16 kHz.
\begin{itemize}
    \item \textbf{S-100hr.} In our setting, we limited the total duration of speech data to 100 hours. Instead of directly using the \textit{train-clean-100} split in LibriSpeech, we select utterances from the \textit{train-clean-100}, \textit{train-clean-360}, and \textit{train-other-500} splits, making a new split with clean and noisy speech data. This split contains 29721 speech clips from 245 speakers (123 males and 122 females) without text transcriptions.
    \item \textbf{ST-10hr.} We use the 10-hour split of Libri-light~\cite{kahn2020libri} as the low resource paired dataset. This split contains 2763 utterances from 24 speakers, and the total duration is about 10 hours. Speech utterances in \textit{ST-10hr} are included in \textit{S-100hr}.
    \item \textbf{S-960hr.} This split includes all training splits in LibriSpeech, which contains 281241 utterances from 2338 speakers with a total duration of about 960 hours. We use the split to pre-train the topline HuBERT model in our experiments.
    \item \textbf{T-960hr.} This split includes all text transcriptions of the total 960 hours data in LibriSpeech. These transcriptions are used to generate the synthetic dataset.
    \item \textbf{T-LM.} This split includes the text used to train the language model in LibriSpeech. The data is used to generate the large synthetic dataset in Section~\ref{ssec:res-s4}.
    \item \textbf{dev-clean} and \textbf{dev-other.} We use the \textit{dev-clean} and \textit{dev-other} splits of LibriSpeech for validation while building the systems and for evaluating the performance.
\end{itemize}

We use the MUSAN dataset~\cite{snyder2015musan} for data augmentation when training the SE model and augmenting the synthetic dataset. We replaced the \textit{speech} part of MUSAN with S-100hr to prevent using additional speech data.

\vspace{-0pt}
\subsection{Comparing Systems}
\label{ssec:compare}
\begin{itemize}
    \item \textbf{Baseline HuBERT (S0).} The baseline in this work is a HuBERT BASE model pre-trained only on limited speech data, \textit{S-100hr}. We pre-train the model for three iterations. Models in different iterations are pre-trained for 200k steps, 400k steps, and 400k steps, respectively. After pre-training, the model is fine-tuned for the ASR task on \textit{ST-10hr} for 40k steps. In each iteration, we experiment with different settings and select the best model as the teacher~\cite{hsu2021hubert} for the next iteration, as shown in Table~\ref{tab:s0-wer}. In this paper, we denote different S0 models as \textit{S0-$x^{th}$-km$y$-last (best)-l$z$}, which means the model is from the $x^{th}$ iteration, the cluster number for k-means is set to $y$, the last (best) checkpoint during pre-training is selected, and we are referring to the $z^{th}$ layer of the model.
    
    We use the official HuBERT implementation and configurations provided in the Fairseq toolkit~\cite{ott2019fairseq} for pre-training and fine-tuning. The model is pre-trained on 32 GPUs and fine-tuned on 8 GPUs. If not specified, the HuBERT models in the following systems are pre-trained and fine-tuned with the same setting of \textit{S0-1st-km100}.
    \item \textbf{Topline HuBERT (S1).} Similar to the BASE model in \cite{hsu2021hubert}, S1 is pre-trained on \textit{S-960hr}. This model is considered a topline system built with high-resource speech data.
    \item \textbf{Off-the-shelf TTS (S2 and S3).} To show the performance that can be reached with a high-quality TTS model trained on the high-resource TTS dataset, we select two off-the-shelf TTS models released in ESPnet~\cite{hayashi2020espnet}. In S2 and S3, we adopt VITS models~\cite{kim2021conditional} trained on the VCTK dataset~\cite{yamagishi2019vctk} and on the clean splits (\textit{train-clean-100} and \textit{train-clean-360}) of LibriTTS~\cite{zen2019libritts}, respectively. We use the pre-trained TTS models and \textit{T-960hr} to generate two synthetic datasets. While generating, the target speakers are randomly sampled from the seen speakers. There are 108 speakers in S2 and 1151 in S3. Two HuBERT models are then pre-trained on the different synthetic datasets, respectively.
    \item \textbf{Proposed (S4).} To build the proposed system, we first extract discrete units from the speech data in \textit{S-100hr} and \textit{ST-10hr}. We use the features from the best \textit{S0-3rd} model (refer to Section~\ref{ssec:res-s0s1}) and generate units as described in Section~\ref{ssec:s2u}. T2U is trained on limited paired data, \textit{ST-10hr}, while SE, DP, PP, and U2S are trained on limited speech data, \textit{S-100hr}. All modules are trained on 8 NVIDIA V100 GPUs, except for SE on 1 NVIDIA 2080Ti GPU. We use \textit{T-960hr} to generate the dataset. Target speakers are randomly selected from the 245 speakers in \textit{S-100hr}. The synthetic dataset is then augmented and combined with \textit{S-100hr} for HuBERT pre-training.
\end{itemize}

\vspace{-0pt}
\subsection{Evaluation Methods}
To evaluate the quality of the discrete units extracted using S0 in different iterations, we calculate phone purity (PP) and cluster purity (CP)~\cite{hsu2021hubert} on \textit{dev-clean} and \textit{dev-other}. For the fine-tuning results, we report the word error rate (WER) on \textit{dev-other}, decoded with a 4-gram language model trained on the official Librispeech language modeling data. All the evaluation metrics are from the official HuBERT implementation in the Fairseq toolkit.

\vspace{-0pt}
\section{Results}
\label{sec:res}

\vspace{-0pt}
\subsection{Performance without Synthetic Data}
\label{ssec:res-s0s1}
\begin{table}[t]
  \vspace{-0pt}
  \caption{Results of pre-training on natural speech data.}
  \label{tab:s0-wer}
  \vspace{-0pt}
  \centering
  \resizebox{0.48\textwidth}{!}{
  \begin{tabular}{cccc}
  \toprule
  \textbf{Feat.} & \textbf{K} & \textbf{WER-last (\%)} & \textbf{WER-best (\%)}  \\ \hline \hline
  \multicolumn{4}{c}{\textbf{\textit{S0-1st}}} \\
  mfcc & 100 & \textbf{25.0} / 25.1 & 31.1 / 33.8 \\ \hline \hline 
  \multicolumn{4}{c}{\textbf{\textit{S0-2nd}}} \\
  S0-1st-km100-last-l6 & 100 & 33.6 / 32.4 & 27.6 / \textbf{26.7} \\ 
  S0-1st-km100-last-l6 & 500 & 38.2 / 37.8 & 27.1 / 27.6 \\ \hline \hline
  \multicolumn{4}{c}{\textbf{\textit{S0-3rd}}} \\
  S0-2nd-km100-best-l9 & 100 & 34.1 & 26.5 \\
  S0-2nd-km100-best-l9 & 500 & 34.6 & \textbf{25.4} \\ \hline \hline 
  \multicolumn{4}{c}{\textbf{\textit{S1-1st}}} \\
  mfcc & 100 & \textbf{14.2} & 14.4 \\ \bottomrule
  \end{tabular}}
  \vspace{-0pt}
\end{table}
\begin{table}[t]
  \caption{Quality analysis of discrete units.}
  \label{tab:s0-pur}
  \vspace{-0pt}
  \centering
  \resizebox{0.3\textwidth}{!}{
  \begin{tabular}{ccccc}
  \toprule
  \textbf{Layer} & \textbf{K} & \textbf{PP (\%)} & \textbf{CP (\%)}  \\ \hline \hline
  \multicolumn{4}{c}{\textbf{\textit{S0-1st-km100-last}}} \\
  6 & 100 & 48.36 & 21.20 \\
  6 & 500 & 58.21 & 8.59 \\ \hline \hline
  \multicolumn{4}{c}{\textbf{\textit{S0-2nd-km100-best}}} \\
  9 & 100 & 54.69 & 26.25 \\
  9 & 500 & 64.32 & 9.42 \\ \hline \hline
  \multicolumn{4}{c}{\textbf{\textit{S0-3rd-km500-best}}} \\
  6 & 100 & 55.43 & 23.87 \\
  9 & 100 & 59.03 & 27.53 \\
  11 & 100 & 58.38 & 27.22 \\
  12 & 100 & \textbf{60.49} & \textbf{28.03} \\
  6 & 500 & 61.93 & 7.82 \\
  9 & 500 & \textbf{67.32} & 9.39 \\
  11 & 500 & 67.05 & 9.54 \\
  12 & 500 & 66.97 & \textbf{9.57} \\ \hline \hline 
  \multicolumn{4}{c}{\textbf{\textit{Pre-trained HuBERT BASE}}} \\
  6 & 500 & 68.12 & 8.99 \\ \bottomrule
  \end{tabular}}
  \vspace{-0pt}
\end{table}
\begin{table}[t]
  \caption{Results of pre-training on synthetic speech data generated by off-the-shelf TTS methods.}
  \label{tab:s2s3-wer}
  \vspace{-0pt}
  \centering
  \resizebox{0.48\textwidth}{!}{
  \begin{tabular}{cccccc}
  \toprule
  \textbf{Sys.} & \textbf{\# of spk.} & \textbf{Nat. (hr)} & \textbf{Synth. (hr)} & \textbf{WER-last (\%)} & \textbf{WER-best (\%)}  \\ \hline \hline
  S0 & 245 & 100 & - & 25.0 & 31.1 \\
  S1 & 2338 & 960 & - & 14.2 & 14.4 \\ \hline
  S2 & 108 & - & 652 & 34.3 & 50.0 \\
  S3 & 24 & - & 800 & 27.3 & 27.0 \\
  S3 & 245 & - & 805 & 26.0 & 26.0 \\
  S3 & 1151 & - & 809 & 23.2 & 23.2 \\
  S3 & 2338 & - & 826 & 22.3 & \textbf{22.2} \\ \bottomrule
  \end{tabular}}
  \vspace{-0pt}
\end{table}
\begin{table*}[t]
  \vspace{-0pt}
  \caption{Results of pre-training on synthetic speech data generated by the proposed method.}
  \label{tab:s4-wer}
  \vspace{-0pt}
  \centering
  \resizebox{0.95\textwidth}{!}{
  \begin{tabular}{cccccccccc}
  \toprule
  \textbf{Sys.} & \textbf{\# of spk.} & \textbf{Spk. per uttr.} & \textbf{Nat. (hr)} & \textbf{Synth. (hr)} & \textbf{Oversampling rate} & \textbf{Steps} & \textbf{Iter.} & \textbf{Synth. FT (hr)} & \textbf{WER (\%)} \\ \hline \hline
  \multicolumn{10}{c}{\textbf{\textit{Fine-tuning on ST-10hr}}} \\
  S0 & 245 & - & 100 & - & - & 200k & 1 & - & 25.0 \\
  S0 & 245 & - & 100 & - & - & 400k & 2 & - & 26.7 \\
  S0n & 245 & - & 100 & - & - & 200k & 1 & - & 26.7 \\
  S1 & 2338 & - & 960 & - & - & 200k & 1 & - & \textbf{14.2} \\
  S2 & 108 & 1 & - & 652 & - & 200k & 1 & - & 34.3 \\
  S3 & 1151 & 1 & - & 809 & - & 200k & 1 & - & 23.2 \\ \hline
  S4 & 245 & 1 & - & 1.1k & - & 200k & 1 & - & 23.5 \\
  S4 & 245 & 1 & 100 & 1.1k & 1 & 200k & 1 & - & 21.5 \\
  S4 & 245 & 1 & 100 & 1.1k & 1 & 400k & 1 & - & 22.2 \\
  S4 & 245 & 10 & 100 & 11k & 1 & 200k & 1 & - & 23.8 \\
  S4 & 245 & 10 & 100 & 11k & 10 & 200k & 1 & - & 21.5 \\
  S4 & 245 & 10 & 100 & 11k & 10 & 400k & 1 & - & 21.0 \\
  S4 & 245 & 10 & 100 & 11k & 100 & 200k & 1 & - & \textbf{20.4} \\ \hline \hline
  \multicolumn{10}{c}{\textbf{\textit{Fine-tuning on ST-10hr and synthetic data}}} \\
  S4 & 245 & 10 & 100 & 11k & 100 & 200k & 1 & 100 & 19.2 \\
  S4 & 245 & 10 & 100 & 11k & 100 & 200k & 1 & 1k & 17.9 \\
  S4 & 245 & 10 & 100 & 11k & 100 & 200k & 1 & 10k & \textbf{17.5} \\
  S4 & 245 & 1 & 100 & 1.1k & 100 & 200k & 2 & 10k & \textbf{15.8} \\
  S4 & 245 & 10 & 100 & 11k & 100 & 200k & 2 & 10k & 17.2 \\ \bottomrule
  \end{tabular}}
  \vspace{-0pt}
\end{table*}

We first evaluate the performance of the baseline model, S0, pre-trained on mere 100 hours of speech data of \textit{S-100hr}. The results are presented in Table~\ref{tab:s0-wer}. The column \textbf{Feat.} indicates the type of the pre-training target, which is either mfcc or representations extracted from the model of the previous iteration with the best WER result. \textbf{K} denotes the number of clusters for k-means clustering. We report the WERs on the \textit{dev-other} split of LibriSpeech. \textbf{-last} and \textbf{-best} represent that the last and the best model checkpoints during pre-training are used for fine-tuning, respectively. It is worth noting that the \textbf{WER-last} gets worse in the $2^{nd}$ and $3^{rd}$ iterations, deviating from the findings in \cite{hsu2021hubert}. This divergence could be attributed to overfitting as we constrain the amount of training data to 100 hours only. Even fine-tuning with the best checkpoints during pre-training fails to enhance the performance. Thus, we can deduce that a HuBERT BASE model can only reach a WER of approximately 25\% when pre-trained on limited speech data. For comparison, the topline model (S1) attains 14.2\% when pre-trained on 960 hours of speech data.

As the proposed system relies on discrete units generated using S0 and k-means clustering, we first evaluate the quality of the units. We select S0 models in different iterations with the best WERs, which are \textit{S0-1st-km100-last}, \textit{S0-2nd-km100-best}, and \textit{S0-3rd-km500-best}. We also consider the officially released pre-trained HuBERT BASE model for comparison. The features of different layers are extracted from \textit{S-100hr}, then k-means clustering with different $k$ is conducted on these features. The results are listed in Table~\ref{tab:s0-pur}. Unlike our previous findings, the quality of units improves with more pre-training iterations. We select \textit{S0-3rd-km500-best-l9} and $k=500$ to generate units for the proposed system as the setting is the most performant.

\vspace{-0pt}
\subsection{Performance of Off-the-Shelf TTS Methods}
\label{ssec:res-s2s3}
We use off-the-shelf TTS methods to generate synthetic datasets and pre-train a HuBERT models, as described in Section~\ref{ssec:compare}. The results are listed in Table~\ref{tab:s2s3-wer}. \textbf{Nat.} and \textbf{Synth.} denote the total duration of the natural and the synthetic data for pre-training, respectively. The dataset generated by S2 is smaller and contains only 652 hours of speech data. The WERs are also much worse than the baseline. For S3, the TTS model is trained on a dataset with 1151 speakers and adopts an x-vector to specify the target speaker. When synthesizing a dataset, we randomly sample x-vectors of 24, 245, and all 1151 speakers from the training data and all 2338 speakers from \textit{S-960hr}. The results show that speaker diversity of the dataset is crucial for pre-training a good HuBERT model. With the number of speakers increasing from 24 (same as \textit{ST-10hr}), 245 (same as \textit{S-100hr}), to 1151, the WER improves accordingly from 27.0\%, 26.0\%, to 23.2\%. The best result of S3 is about 22.2\%.

\vspace{-0pt}
\subsection{Performance of the Proposed System}
\label{ssec:res-s4}
Table~\ref{tab:s4-wer} shows the performance of the proposed system. We report \textbf{WER-last} except for the $2^{nd}$ iteration of S0. \textbf{Spk. per uttr.} indicates how many speech utterances of different speakers are generated for each text utterance. \textbf{Synth. FT} denotes the size of additional synthetic data used for fine-tuning. We conclude our observations as follows: (1) augmenting the 100 hours natural speech with background noise, denoted as S0n, does not improve the performance; (2) the proposed TTS method attains a similar WER to S3 (23.5\% vs. 23.2\%), while the latter uses a large high-quality dataset to build the TTS model; (3) generating the same utterance with different speaker styles and a higher oversampling rate for the natural speech effectively improve the WER to 20.4\%; (4) increasing the pre-training steps from 200k to 400k slightly helps for a larger synthetic dataset.

To further improve the performance, we augment the fine-tuning data with the paired data synthesized from \textit{T-LM}. Note that this text dataset has been used in all systems to build the 4-gram language model for decoding. We generate extra 100, 1k, and 10k hours of paired data for fine-tuning and improve the WER by 3\% to 17.5\%. Lastly, we run the pre-training for the $2^{nd}$ iteration. The best WER attains 15.8\%, which is close to the performance of the topline. Compared with S0, the proposed system decreases the WER by about 10\% (from 25.0\% to 15.8\%). Compared with S1, the proposed system reduces the demand for speech data by approximately 90\% (from 960 hours to 100 hours) with only slight degradation in WER.

\vspace{-0pt}
\section{Conclusion}
\label{sec:con}
This paper proposed an iterative process to build a TTS model and a HuBERT model with extremely limited data. By using synthetic data for pre-training, the proposed method drastically alleviate the demand for large amounts of real-world speech data required for pre-training an SSL model.

\bibliographystyle{IEEEtran}
\bibliography{mybib.bib}
\end{document}